\documentclass[aps,jcp,preprint,showpacs,groupedaddress]{revtex4-1}

\usepackage[intlimits]{amsmath}
\usepackage[dvips]{graphicx}
\usepackage{amsfonts}
\usepackage{amssymb}
\usepackage{graphicx}
\usepackage{bm} 
\usepackage{natbib}
\usepackage{color}
\usepackage{subfig}

\begin{document}

%\begin{titlepage}

\title{Theoretical analysis of degradation mechanisms in the formation of morphogen gradients}

\author {Behnaz Bozorgui, Hamid Teimouri, and Anatoly B. Kolomeisky}
\affiliation{Department of Chemistry and Center for Theoretical Biological Physics, Rice University, Houston, TX 77005-1892}

\begin{abstract}

The fundamental biological processes of development of tissues and organs in multicellular organisms is governed by various signaling molecules, which are called morphogens. It is known that spatial and temporal variations in concentration profiles of signaling molecules, which are frequently referred as morphogen gradients,  lead to  cell differentiation via activating specific genes in a concentration-dependent manner. It is widely accepted that the establishment of the morphogen gradients involves  multiple biochemical reactions and diffusion processes. One of the critical elements in the formation of  morphogen gradients is a degradation of signaling molecules. We develop a new theoretical approach that provides a comprehensive description of the degradation mechanisms. It is based on the idea that the degradation works as an effective potential that drives the signaling molecules away from the source region. Utilizing the method of first-passage processes, the dynamics of the formation of morphogen gradients for various degradation mechanisms is explicitly evaluated. It is found that the linear degradation leads to a dynamic behavior specified by times to form the morphogen gradients that depend linearly on the distance from the source. This is  because the effective potential due to degradation is quite strong.  At the same time, the nonlinear degradation mechanisms yield a quadratic scaling in the morphogen gradients formation times since the effective potentials are much weaker. Physical-chemical explanations of these phenomena are presented.

\end{abstract}

\maketitle

%\end{titlepage}

\section{Introduction}

The development of multi-cellular organisms is one of the most important fundamental processes in nature \cite{Martinez_book,lodish_book,wolpert_book}. The most critical question here is how a small set of genetically identical cells in embryos can produce morphologically and functionally different tissues and organs in fully developed organisms. The central concept of biological development is that the observed complex spatial patterning is a result of action of signaling molecules that are also called morphogens \cite{Martinez_book,lodish_book,wolpert_book,wolpert69,tabata04,lander07,porcher10,rogers11}. Signaling molecules can produce non-uniform concentration profiles, the so-called morphogen gradients, that via complex biochemical networks stimulate or suppress specific genes in embryo cells, depending on the local concentration.  In recent years, there were multiple experimental and theoretical investigations on how the morphogen gradients are created and how they function. This led to several exciting discoveries in the field \cite{lander07,rogers11,tabata04,porcher10,gregor07,kicheva07,zhou12,yu09,entchev00,muller12,kerszberg98,spirov09,drocco11,little11,fedotov14,berezhkovskii10,gordon13,kolomeisky11,houchmandzadeh02,england05}. However, many aspects of the mechanisms for formation of the morphogen gradients remain not fully explained \cite{kornberg12}.

A large variety of approaches to describe the development of morphogen gradient have been proposed and discussed \cite{lander07,porcher10,kornberg12}. Many of them follow the original idea of Turing that the morphogen gradients are resulting from complex reaction-diffusion process \cite{turing52}. The most popular and widely utilized method to explain the formation of signaling molecules profiles is known as a synthesis-diffusion-degradation (SDD) model \cite{porcher10,kicheva07,drocco12}. In this picture, the  process starts with morphogens being produced at specific localized regions in the embryo, from which they diffuse along the cells. Signaling molecules also can be removed from the system after binding to specific receptors on cells.  At large times, this leads to exponential decaying concentration profiles which qualitatively agree with many experimentally observed morphogen gradients \cite{porcher10,gregor07,kicheva07,zhou12,yu09,grimm09,drocco12}.    

It is widely accepted that the process of degradation or removal of signaling molecules from the system is critically important for the development of morphogen gradients \cite{porcher10}. This allows the formation of the stationary profiles of signaling molecules, ensuring the robustness of the  genetic information transfer in biological development. But specific details of how the degradation influences the formation  of morphogen gradients are still not well clarified. There are many counter-intuitive observations that cannot be explained by current theoretical views. In the classical SDD model it is assumed that the degradation is linear, i.e., the particle flux leaving the system is proportional to the local  concentration of morphogens. It was shown theoretically then that for this model the time to establish  a stationary morphogen gradient at given location, which is also known as a local accumulation time (LAT), is a linear function of the distance from the source \cite{berezhkovskii10}. This observation is surprising since for the system with unbiased diffusion of particles much more slower quadratic scaling was expected \cite{berezhkovskii10,kolomeisky11}. At the same time, several experiments suggested that in some cases the establishment of morphogen gradients is associated with nonlinear degradation mechanisms when the presence of signaling molecules self-enhances or self-catalyzes its removal from the system  \cite{eldar03,gordon11,chen96,incardona00}.  Theoretical investigations of temporal evolution of the morphogen gradients with nonlinear degradation suggested that in this case the local accumulation times, in contrast to linear degradation, scale quadraticaly with the distance from the source \cite{gordon11}. But the presented mathematical analysis was rather very complicated, and only bounds for LAT in several cases where obtained \cite{gordon11}.   

These observations raised several interesting and important questions concerning the role of the degradation in regulating the concentration profiles of signaling molecules. Why the degradation accelerates the relaxation to the stationary state for linear degradation? Why the actions of linear and nonlinear degradation processes are so different? What is the physical mechanism of degradation? Recently, one of us proposed an idea that might resolve some of these issues \cite{kolomeisky11}. It was suggested that the degradation acts as an effective potential that pushes signaling molecules away from the source region. It means that the degradation will make the diffusion of morphogen molecules effectively biased.  However, only qualitative arguments have been presented.     

In this paper, we extend  and generalize the original idea that the removal of signaling molecules works as the effective potential.  A new quantitative approach that provides a microscopic view on the role of degradation in the formation of morphogen gradients is developed. It allows us to explain the differences between various degradation mechanisms. We argue that the linear degradation corresponds to a strong potential, leading to  strongly biased motion of the signaling molecules. At the same time, the non-linear degradation creates a potential that is too weak to modify the underlying random-walk scaling behavior of the system, affecting only the magnitude of fluctuations.

\section{Theoretical Method}

Let us start the analysis of degradation mechanisms by introducing a discrete SDD model as presented in Fig. 1a. The cells in the embryo are represented as discrete sites $n \ge 0$ on this semi-infinite lattice. The signaling molecules are produced at the origin ($n=0$) with a rate $Q$. Then morphogens diffuse along the lattice with a diffusion constant $D$. At each lattice site $n$ the molecule can be degraded with a rate $k_{n}$. It is convenient to adopt a single-molecule view of the process where the local concentration of signaling molecules is proportional to a probability to find the morphogen molecule at a given location \cite{kolomeisky11,teimouri14}. One can define then  $P_n(t)$ as a probability of finding the morphogen at the site $n$ at time $t$.  These probabilities evolve with time as described by a set of master equations, 
\begin{equation}\label{eq:master1}
  \frac{dP_n(t)}{dt} = DP_{n+1}(t) + DP_{n-1}(t)-(2D+k_n)P_n(t),  
\end{equation}
for $n>0$; while at the origin ($n=0$) we have
\begin{equation}\label{eq:master2}
  \frac{dP_0(t)}{dt} = Q+ DP_{1}(t)- (D+k_{0})P_0(t).
\end{equation}

The situation when the degradation rate $k_{n}$ is independent of the concentration of signaling molecules corresponds to  linear degradation since the total flux that removes morphogens from the system [$k_{n} P_{n}(t)$] is proportional to the concentration. For the case of constant $k_{n}=k$, this discrete SDD model with linear degradation was fully analyzed before \cite{kolomeisky11}. In a more general scenario, the degradation rate might depend on the local concentration, $k_n=kP_n^{m-1}$, where a parameter $m$ specifies the degree of non-linearity, and this corresponds to non-linear degradation processes. However, it is not feasible generally to obtain full analytic solutions for these non-linear degradation models (with $m>1$).    

The main idea of our approach is that degradation acts as an effective potential. This suggests that the original reaction-diffusion process with degradation is equivalent to a biased diffusion process in such potential but {\it without} degradation, as shown in Fig. 1b. To explain the origin of this potential, let us consider the system in the steady-state limit when a stationary non-uniform profile $P_{n}^{s}$ is achieved. The degradation leads to a concentration gradient between any two consecutive sites, and this gradient can be associated with a difference in the chemical potentials of the morphogens,
\begin{equation}
\mu_{n+1} -\mu_n = k_{B}T\ln{P^{(s)}_{n+1}} -k_{B}T\ln P^{(s)}_n.
\end{equation} 
This can also be viewed as an effective potential that influences particles that are not degraded. It follows then that this potential can be evaluated as  
\begin{equation}
U_{n}^{eff} = k_{\textrm B}T\ln{P^s_{n}}.
\end{equation}

The above arguments indicate that dynamics of the reaction-diffusion model (Fig. 1a) can be mapped into  the biased-diffusion model (see Fig. 1b), which is much simpler to analyze. For the equivalent biased diffusion model we define $\Pi_{n}(t)$ as the probability of finding a particle at position $n$ at time $t$. These probabilities are also governed  by corresponding master equations,
\begin{equation}\label{eq:master3}
  \frac{d\Pi_n(t)}{dt} = r_{n+1}\Pi_{n+1}(t) + g_{n-1}\Pi_{n-1}(t)-(r_{n}+g_{n})\Pi_n(t),  
\end{equation}
for $n>0$; and 
\begin{equation}\label{eq:master4}
  \frac{d\Pi_0(t)}{dt} = Q+ r_{1}\Pi_{1}(t)- g_{0}\Pi_0(t),
\end{equation}
for $n=0$. The diffusion rates $g_{n}$ and $r_{n}$ are related to each other via the effective potential as can be shown using the detailed balance arguments \cite{vankampen}: 
\begin{equation}\label{eq:DB}
  \frac{g_n}{r_{n+1}}=\exp \left (  \frac{U_{n}^{eff}-U_{n+1}^{eff}}{k_{\textrm B}T} \right)
\end{equation}
This is an important result because it directly couples the original SDD model with degradation to the new biased-diffusion model without degradation.

One more step is needed in order to have comparable dynamic behaviors in both models. The average residence times for the particles at each site provide a measure of relevant time scales in the system. It seems reasonable to require that these quantities to be the same in both models, leading to   
\begin{equation}\label{eq:res}
  g_n+r_n=2D+k_n. 
\end{equation}
Note that Eqs. \ref{eq:DB} and \ref{eq:res} uniquely define forward and backward rates in the biased-diffusion model.

To understand the mechanisms of formation of the morphogen gradients the relaxation dynamics to a stationary-state behavior needs to be investigated. This can be done by analyzing the local accumulation times $t_n$, which are defined as times to reach the stationary state concentration at given position $n$. The general approach for computing LAT is known \cite{berezhkovskii10}, but analytical results can only be obtained for the linear degradation model ($m=1$). We propose to use mean first-passage times $\tau_n$ (MFPT), which are defined as times to reach a given site for the first time,  as a measure of dynamics of the formation of morphogen gradients.  It was shown before that  MFPT approximate very well LAT at large distances from the source, i.e., for large $n$ \cite{kolomeisky11,berezhkovskii11}. In addition, the first-passage analysis provides more clear physical view of the underlying phenomena in the development of morphogen gradients. 

Thus, our method of evaluating the formation of signaling molecules profiles consists of three steps. First, from the original SDD model with degradation the stationary-state profiles are obtained, from which the effective potentials are explicitly evaluated. In the second step, the transition rates in the equivalent biased-diffusion model without degradation are computed. Finally, these rates are utilized for calculating the first-passage dynamics in the system. It is important to note here that this procedure is not exact since it involves several approximations.

\section{Linear degradation}

To test our theoretical approach, we start with the simplest linear degradation model where all dynamic properties are analytically calculated for all sets of parameters \cite{berezhkovskii10,kolomeisky11}. The stationary-state profile for the SDD model can be easily evaluated \cite{kolomeisky11},
\begin{equation}\label{ld:profile}
P_n^{(s)}=\frac{2Q x^n}{k+\sqrt{k^{2}+4Dk}},
\end{equation}
with $x= (2D+k-\sqrt{k^2+4kD})/2D$.  This expressions allows us to estimate the effective potential due to degradation for the equivalent biased-diffusion model,
\begin{equation}
\frac{U_{n}^{eff}}{k_{B}T} \simeq n \ln x.
\end{equation}
This potential is linear  with a slope that depends on diffusion and degradation rates. It is also shown in Fig. 5. Employing these results in Eqs. \ref{eq:DB} and \ref{eq:res}, we obtain the following expressions for the forward and backward transition rates,
\begin{equation}
 g_n = g = \frac{2D+k}{x+1}, \quad r_{n+1} = r = x \ \frac{2D+k}{x+1}.
\end{equation}
Note that these rates are independent of the position and  the production rate $Q$.  

In the final step, first-passage dynamics  can be  evaluated by using known expressions for MFPT \cite{vankampen},
\begin{equation}
  \begin{split}	
    \tau_n &= \sum_{i=0}^{n-1}\sum_{j=0}^{i} \frac{r_ir_{i-1}...r_{j+1}}{g_ig_{i-1}...g_{j+1}{g_j}} \\
    &= \frac{(x+1)}{(2D+k)}\frac{\left[x(x^n-1)-n(x-1)\right]}{(x-1)^2}.
  \end{split}
  \label{MFPT}
\end{equation}	
It can be easily checked that in the special case of no degradation in the original system, $k=0$, this formula reduces to $\tau_n \simeq n^{2}/2D$ at large distances, as expected for a simple unbiased random walk. 

It is possible to compare the obtained mean first-passage times from Eq. (\ref{MFPT}) with analytical expressions for LAT and for MFPT in the original SDD model which are available \cite{kolomeisky11}. But it is more convenient first to do it in two different dynamic regimes. In the case when the degradation rate is much faster than diffusion, $k \gg D$, it can be shown that $x \simeq D/k$, which leads to $\tau_{n} \simeq n/k$.  It is in excellent agreement with exact results for LAT and MFPT for the original SDD model in this limit \cite{kolomeisky11}, $t_{n}=\tau_{n}^{SDD} \simeq (n+1)/k$. In the opposite limit of very fast diffusion ($D \gg k$), we have $x \simeq 1-\sqrt{k/D}$ and Eq. (\ref{MFPT}) yields
\begin{equation} 
\tau_{n} \simeq n/\sqrt{kD}.
\end{equation} 
Exact expressions for LAT and MFPT for the original SDD model give us \cite{kolomeisky11},
\begin{equation}
t_{n} \simeq \frac{1}{2k} \left[1+\frac{n+1}{\sqrt{D/k}} \right], \quad \tau_{n}^{SDD} \simeq n/2\sqrt{Dk}. 
\end{equation}
Thus, for large $n$ our method still correctly reproduces the linear scaling in the local accumulation times, but the amplitude deviates in two times. 

The comparison between predicted MFPT for the biased-diffusion model and for LAT of the original SDD model for general sets of parameters is given in Fig. 2. One can see that our method approximates the dynamics of the formation of morphogen gradient reasonably well. The agreement is better for larger  degradation rates where the effective potentials are stronger. At the same time, for weaker degradation rates there are deviations, although the qualitative behavior is correctly captured. This is a remarkable result given how simple is the theory and that it involves several strong  approximations. This also suggests that the method can be reliably applied to more complex systems with non-linear degradation.

\section{Non-linear degradation}

Here we apply our method for systems where the formation of signaling molecules profiles is accompanied by the non-linear degradation with $k_{n}=k P_{n}^{m-1}$ for $m=2,3,...$. To evaluate the effective potential we need to estimate the stationary-state concentration profiles. However, it is not possible to calculate them analytically  for general non-linear discrete SDD models. But we can use the fact that in the continuum limit ($D \gg k_{n}$) the original master equations (\ref{eq:master1}) and (\ref{eq:master2}) can be written as the corresponding non-linear reaction-diffusion equations,
\begin{equation}
\frac{\partial P(n,t)}{\partial t}=D \frac{\partial^{2}P(n,t)}{\partial n^{2}}-kP^{m}(n,t),
\end{equation}
with the boundary condition at the origin
\begin{equation}
D \frac{\partial P}{\partial n} |_{n=0}=-Q.
\end{equation}
These equations can be solved in the steady-state limit, producing 
\begin{equation}
P_{n}^{(s)} \simeq\frac{1}{(1+n/\lambda)^{\frac{2}{m-1}}},
\end{equation}
where the parameter $\lambda$ is given by
\begin{equation}
\lambda=\frac{1}{m-1}\left[ \frac{(2D)^{m}(m+1)}{k Q^{m-1}} \right]^{\frac{1}{m+1}}.
\end{equation}
It can be shown that the continuum $P_{n}^{(s)}$  describes also quite well the stationary-state behavior of the general non-linear discrete SDD models at large distances from the source. This allows us to approximate the effective potentials for non-linear degradation as
\begin{equation}\label{nonlinear_potential}
\frac{U_{n}^{eff}}{k_{B}T} \simeq -\frac{2}{m-1} \ln (1+n/\lambda).
\end{equation}
This potential is logarithmic, and the degree of non-linearity determines its magnitude as illustrated in Fig. 5. It is important to note here that these potentials are {\it always} weaker than the potential for the linear degradation: see Fig. 5.

Now using Eqs. \ref{eq:DB} and \ref{eq:res} one can obtain the expressions for transition rates in the biased-diffusion model,
\begin{equation}
g_{n} = D \left[\frac{P_{n}^{(s)}}{P_{n+1}^{(s)}}\right]^{0.5} , \quad  r_{n+1} = D \left[\frac{P_{n}^{(s)}}{P_{n+1}^{(s)}}\right]^{-0.5}.
\end{equation}
In the final step, again utilizing the analytical framework for the first-passage processes \cite{vankampen}, we derive the explicit expressions for the mean first-passage times that  approximate the formation of morphogen gradients with nonlinear degradation,
\begin{equation}
\tau_{n}=\sum_{i=0}^{n-1}\sum_{j=0}^{i} \frac{r_ir_{i-1}...r_{j+1}}{g_ig_{i-1}...g_{j+1}{g_j}}=\frac{1}{D}\sum_{j=0}^{n-1} [j(j+1)]^{\frac{1}{1-m}} \sum_{l=0}^{j} l^{\frac{2}{m-1}}.
\end{equation}   
It can be shown that this expression asymptotically at large distance  approaches to
\begin{equation}
\tau_n \approx \frac{(m-1)}{(m+1)}\frac{n^2}{2D}.
\end{equation}
This is an important result since it predicts a quadratic scaling for {\it all} non-linear degradation mechanisms with $m>1$. Furthermore, as expected, for very large $m$, which corresponds to effectively no degradation, this formula reduces to a simple random walk dependence.  

Our theoretical estimates for the relaxation dynamics in the establishment of the morphogen gradients for various models with  non-linear degradation are presented in Fig. 3. One can clearly see that the predicted local accumulation times  approach the quadratic scaling for large $n$ for all possible ranges of diffusion and degradation rates. The approach is faster for larger $m$.  The scaling is independent of the degradation mechanisms, and only the amplitude is determined by the degree of the non-linearity $m$.

We also compared theoretical predictions with numerically exact values of LAT for different non-linear degradation models. The results are presented in Fig. 4. A remarkable agreement between predicted and exact relaxation times is found for $m=3$. It can be seen that increasing the strength of the degradation (larger $k$) improves the agreement even for small distances from the sources. For $m=10$ our theory also works qualitatively well, although there are bigger quantitative deviations. It correctly describes the scaling, and increasing the degradation rate $k$ decreases the magnitude of these deviations. 

Analyzing  results given in Figs 3 and 4, we can make several conclusions about the applicability of the developed theoretical method for analyzing nonlinear degradation.  Our approach correctly finds the quadratic scaling in the local accumulation times. It works better for large distances because it calculates only the arrival times which are always smaller than the correct LAT that also must include some local rearrangements. At large distances the contribution from MFPT to LAT becomes dominant \cite{kolomeisky11}. One can also observe that our method works better for stronger degradation, which corresponds to small $m$ values and/or large degradation rates $k$. Most probably, this is due to the fact that our approach neglects particle fluctuations that are present even in the absence of degradation. For strong degradations these fluctuations become less relevant.

\section{Summary and Concluding Remarks}

We developed a new theoretical approach to analyze mechanisms of degradation in the formation of signaling molecules profiles during the biological development. The method is quite simple, and it provides a full analytical description for all ranges of parameters. It is based on the idea that degradation is similar to the effective potential imposed to morphogen molecules. The potential pushes signaling molecules away from the source region. It allows us to map the original reaction-diffusion process into the biased-diffusion model without degradation, which is much easier to analyze. Finally, utilizing the first-passage approach, the dynamics of relaxation to stationary morphogen gradients can be fully described.   

Despite the fact that our approach involves several strong approximations, it works remarkably well for different models with degradation. We correctly predict the scaling behavior for the local accumulation times in all cases. As we found for both linear and non-linear degradation processes, theoretical method is almost exact for large distances from the source and for faster degradation rates. At the same time, for close distances and for slower degradation rates the agreement is mostly qualitative, although the deviations are relatively small. The effect of the distance can be explained by recalling that in our method first arrival times are computed. The correct LAT involve  local rearrangements which become less important for large distances. The strength and the speed of degradation influence our results because the theoretical method neglects the local particle fluctuations due to underlying random walk dynamics. These fluctuations are expected to contribute significantly to dynamic properties for weak and slow degradations, while they are much less important for strong and fast degradations. 

The advantage of our method is not only the fact that it gives a fully analytical description of the complex processes during the development of the morphogen gradients. It also provides  clear physical explanations for the observed phenomena. We can understand now why linear and non-linear degradation lead to very different dynamic behaviors. For linear degradation we predict that the effective potential is very strong (Fig. 5). The morphogens are strongly pushed away from the source region, and as a result a driven diffusion with the expected linear scaling is observed. For non-linear degradation processes the effective potentials are much weaker (logarithmic versus linear --- see Fig. 5). The particles are moved preferentially in the direction away from the source region, but the underlying random-walk dynamics is not perturbed much. As a result, the quadratic scaling is predicted and the effect of the potential only shows up in the magnitude of fluctuations. These finding also suggest that the degradation might be an effective tool for tuning the complex biochemical and biophysical processes in biological development.

Although the presented method captures main features of the degradation processes during the formation of morphogen gradients, it is important to note that our approach is oversimplified and it involves many approximations. It will be important to test the proposed ideas with more advanced theoretical methods as well as in the extensive experimental studies.

\section*{Acknowledgments}
The work was supported by the Welch Foundation (Grant C-1559), from the NSF (Grant CHE-1360979), and by the Center for Theoretical Biological Physics sponsored by the NSF (Grant PHY-1427654).

\newpage

\noindent Fig.1. a) Schematic view of a discrete synthesis-diffusion-degradation model with unbiased diffusion.   b) Schematic view of equivalent biased-diffusion model without degradation and with modified diffusion rates. Lattice sites correspond to embryo cells.

\vspace{5mm}

\noindent Fig.2. a) Ratio of the calculated mean first passage times and the exact analytical results from the SDD model with linear degradation  as a function of the distance from the source. Different curves correspond to different values of the degradation and diffusion rates. b) The same ratio as a function of the ratio of the  degradation rate over diffusion. Distance from the source is set to $n=10^4$, which exceeds the decay lengths for all values of degradation rates.

\vspace{5mm}

\noindent Fig.3.  Theoretically calculated mean first passage times as a function of the distance from the source for different degrees of non-linearity and  for different values of the degradation rates: a) $m=2$; b) $m=10$.

\vspace{5mm}

\noindent Fig.4.  Theoretically calculated mean first passage times (solid circles) and the numerical exact results from the SDD model (open circles) for the local accumulation times as a function of the distance from the source for different degrees of non-linearity and for different values of degradation rates. For all calculations $D=1$ is assumed.

\vspace{5mm}

\noindent Fig.5.  Effective potentials acting on morphogens due to degradation. Linear degradation corresponds to $m=1$, while $m=3$ and $m=10$ describe different cases of non-linear degradation. For all calculations $k=D=Q=1$ was assumed.

\newpage

\begin{figure}[h]
  \begin{center}
    \unitlength 1in  
      \subfloat[Synthesis-diffusion-degradation model]{\label{fig:1a}\resizebox{3.375in}{1in}{\includegraphics{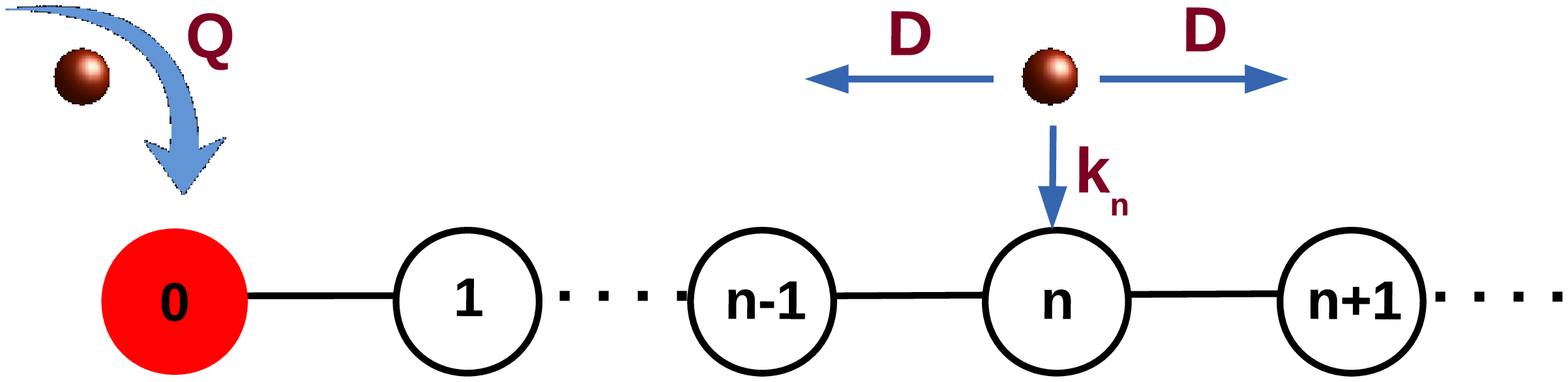}}}\\
      \subfloat[Biased-diffusion model]{\label{fig:1b}\resizebox{3.375in}{1in}{\includegraphics{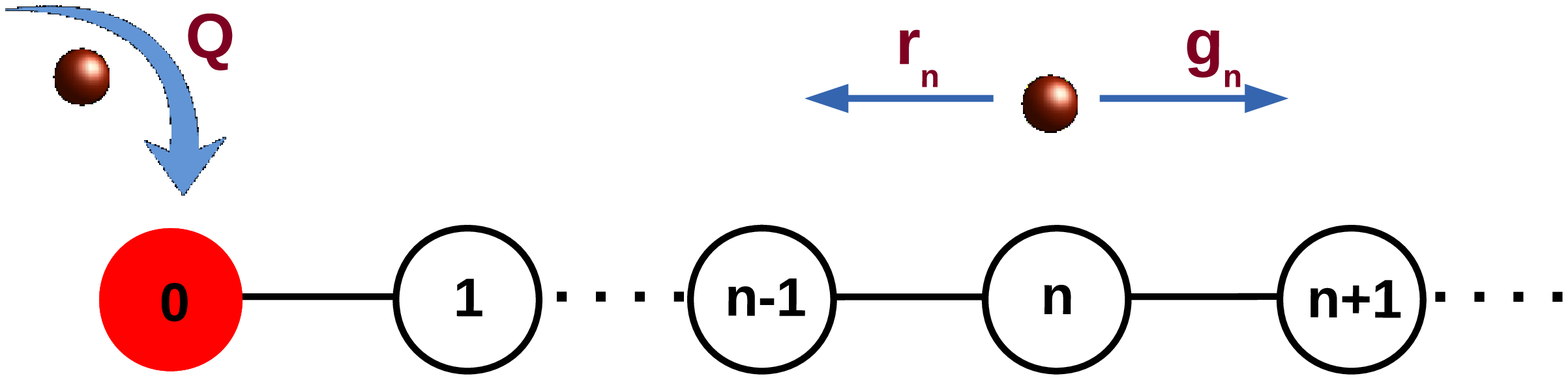}}}
    \vskip 1in
    \begin{Large} Figure 1. Bozorgui, Teimouri, and Kolomeisky \end{Large}
    \caption{}
  \end{center}
\end{figure}

\newpage

\begin{figure}[h]
  \begin{center}
    \unitlength 1in  
    \subfloat[]{\label{fig:2a}\resizebox{3.375in}{2.5in}{\includegraphics{fig2a.eps}}}\\
    \subfloat[]{\label{fig:2b}\resizebox{3.375in}{2.5in}{\includegraphics{fig2b.eps}}}
    \vskip 1in
    \begin{Large} Figure 2. Bozorgui, Teimouri, and Kolomeisky \end{Large}
    \caption{}
  \end{center}
\end{figure}

\newpage

\begin{figure}[h]
  \begin{center}
    \unitlength 1in  
    \subfloat[m=2]{\label{fig:3a}\resizebox{3.375in}{2.5in}{\includegraphics{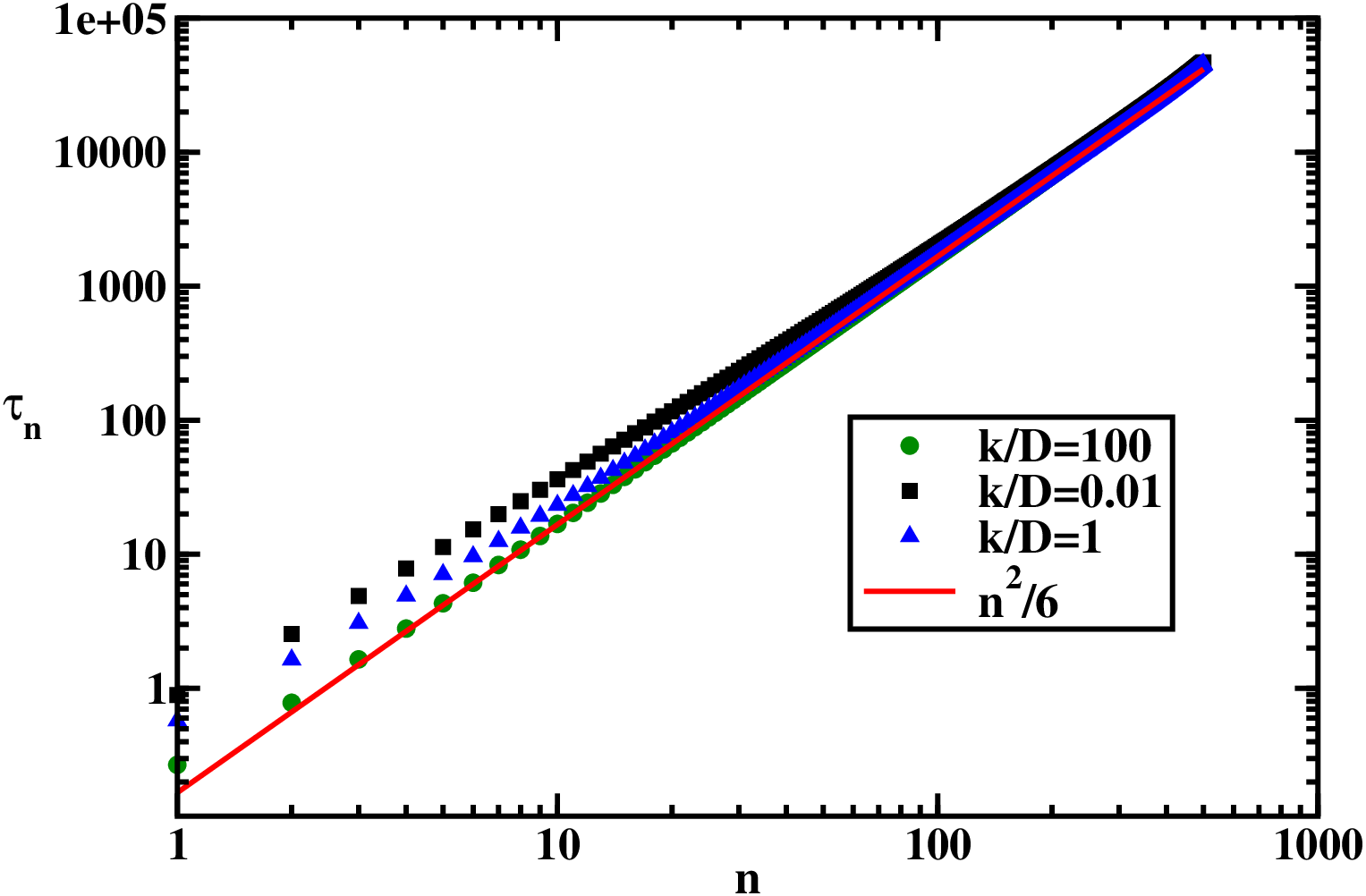}}}
    \subfloat[m=10]{\label{fig:3b}\resizebox{3.375in}{2.5in}{\includegraphics{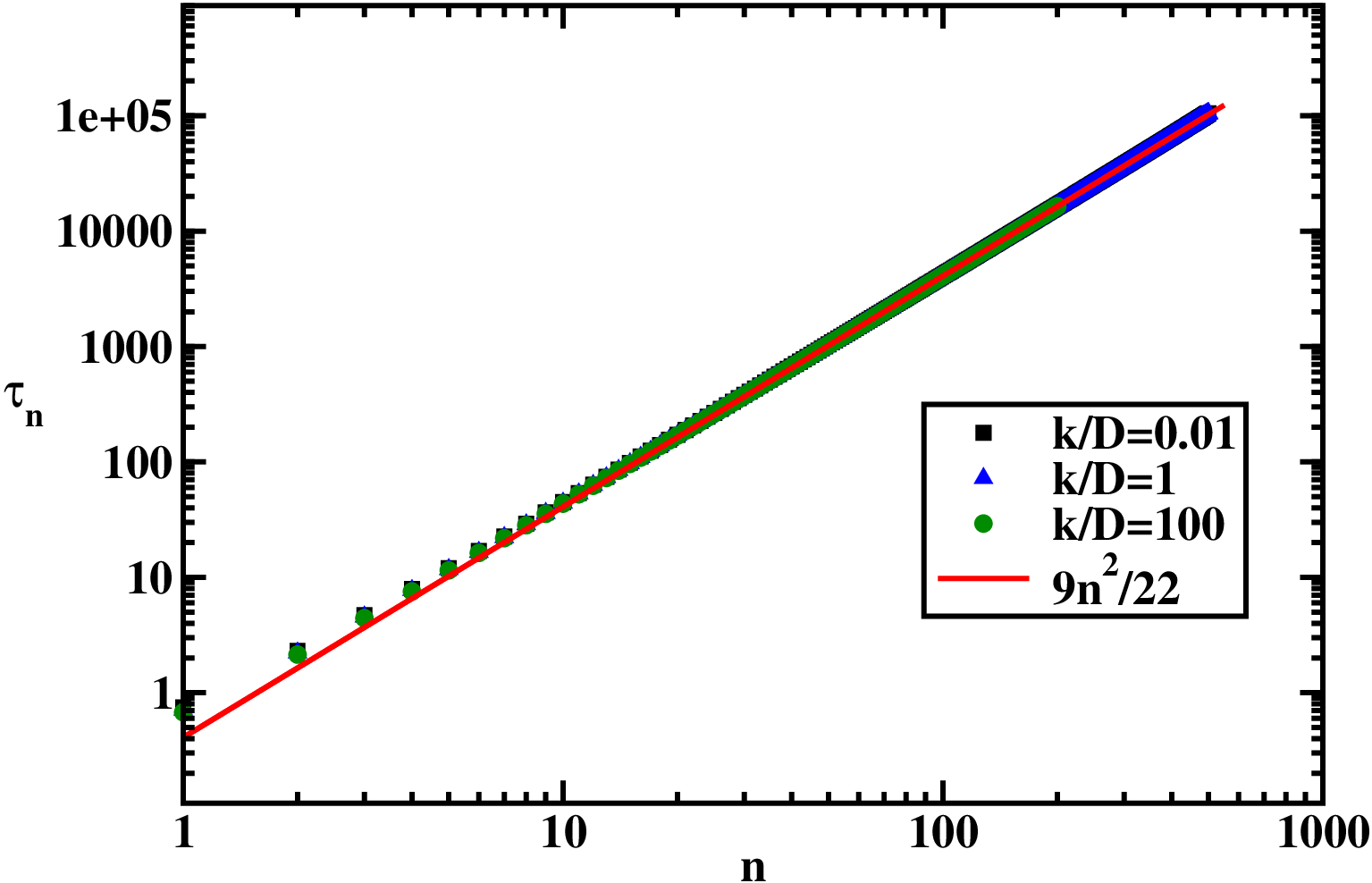}}}
    \vskip 1in
    \begin{Large} Figure 3. Bozorgui, Teimouri, and Kolomeisky \end{Large}
    \caption{}
    \label{fig:3}
  \end{center}
\end{figure}

\newpage

\begin{figure}[h]
  \begin{center}
    \unitlength 1in  
    \label{fig:4a}\resizebox{3.375in}{3.0in}{\includegraphics{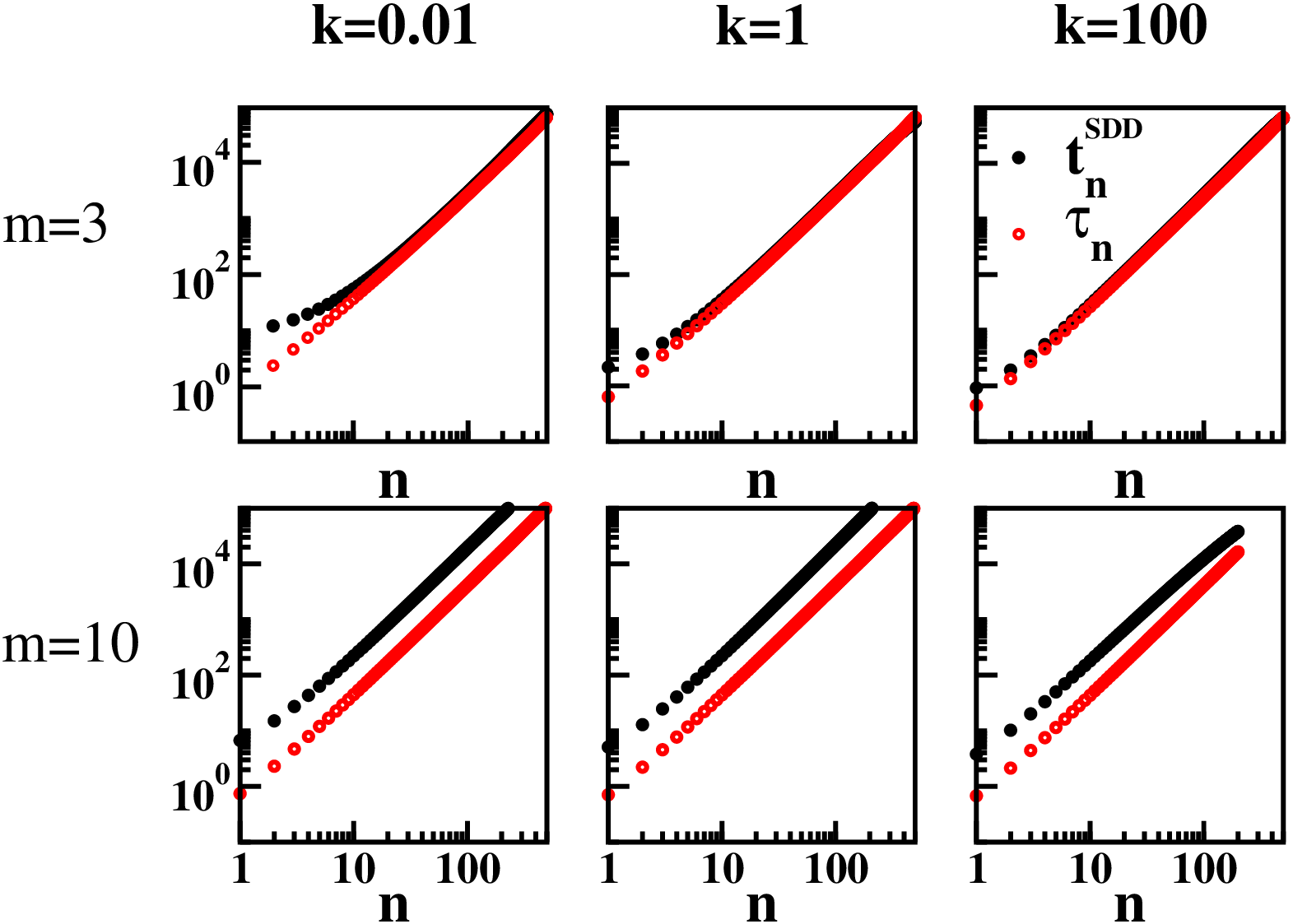}}
    \vskip 1in
    \begin{Large} Figure 4. Bozorgui, Teimouri, and Kolomeisky \end{Large}
    \caption{}
    \label{fig:4}
  \end{center}
\end{figure}

\newpage

\begin{figure}[h]
  \begin{center}
    \unitlength 1in  
    \resizebox{3.375in}{3.0in}{\includegraphics{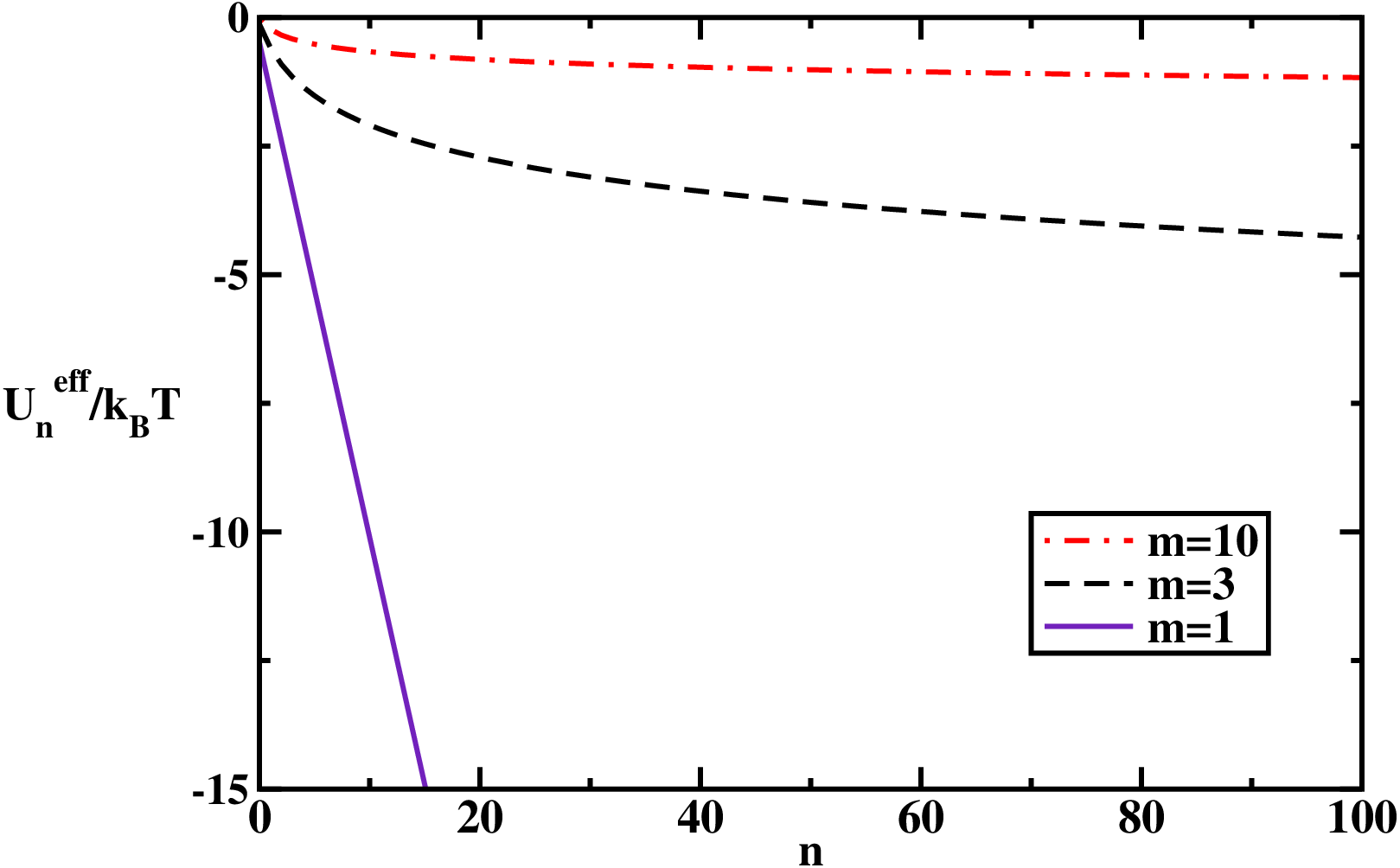}}
    \vskip 1in
    \begin{Large} Figure 5. Bozorgui, Teimouri, and Kolomeisky \end{Large}
    \caption{}
    \label{fig:5}
  \end{center}
\end{figure}

\end{document}